\pdfoutput=1

\documentclass[11pt]{article}

\usepackage{emnlp2021}

\usepackage{times}
\usepackage{latexsym}

\usepackage[T1]{fontenc}

\usepackage[utf8]{inputenc}

\usepackage{microtype}

\usepackage{microtype}
\usepackage{graphicx}
\usepackage{subfigure}
\usepackage{booktabs} 

\usepackage{algorithmic}
\usepackage[ruled,linesnumbered]{algorithm2e}
\usepackage{graphicx}
\usepackage{amsmath}
\usepackage{amssymb}
\usepackage{multirow}
\usepackage{ulem}

\title{Toward Annotator Group Bias in Crowdsourcing}

\author{
	{Haochen Liu\textsuperscript{1}, Joseph Thekinen\textsuperscript{1}\thanks{\hspace{0.2cm}The corresponding author: Joseph Thekinen}, Sinem Mollaoglu\textsuperscript{1}, Da Tang\textsuperscript{2},} \\
	\textbf{Ji Yang\textsuperscript{2}, Youlong Cheng\textsuperscript{2}, Hui Liu\textsuperscript{1}, Jiliang Tang\textsuperscript{1}}
	\vspace{1.6mm}\\
	\fontsize{12}{10}\selectfont\itshape
	\,\textsuperscript{\rm 1}  Michigan State University, East Lansing, MI, USA \\
	\fontsize{12}{10}\selectfont\itshape  \textsuperscript{\rm 2} ByteDance Inc., Mountain View, CA, USA \\
    \fontsize{10}{10}\selectfont \{liuhaoc1,thekinen,sinemm\}@msu.edu;  \{da.tang,ji.yang,youlong.cheng\}@bytedance.com; \{liuhui7,tangjili\}@msu.edu\\}

\begin{document}
\maketitle

\begin{abstract}
Crowdsourcing has emerged as a popular approach for collecting annotated data to train supervised machine learning models. However, annotator bias can lead to defective annotations. Though there are a few works investigating individual annotator bias, the group effects in annotators are largely overlooked. In this work, we reveal that annotators within the same demographic group tend to show consistent group bias in annotation tasks and thus we conduct an initial study on annotator group bias. We first empirically verify the existence of annotator group bias in various real-world crowdsourcing datasets. Then, we develop a novel probabilistic graphical framework \textbf{GroupAnno} to capture annotator group bias with a new extended Expectation Maximization (EM) training algorithm. We conduct experiments on both synthetic and real-world datasets. Experimental results demonstrate the effectiveness of our model in modeling annotator group bias in label aggregation and model learning over competitive baselines.

\end{abstract}

\section{Introduction}

The performance of supervised machine learning algorithms heavily relies on the quality of the annotated training data. Due to the heavy workload of annotation tasks, researchers and practitioners typically take advantage of crowdsourcing platforms to obtain cost-effective annotation data \cite{snow2008cheap,buhrmester2016amazon}. However, the labels collected from multiple crowdsourcing annotators could be not consistent, since the expertise and reliability of the annotators are uncertain, and the task itself could be subjective and difficult. In recent years, a lot of efforts from the machine learning community have been conducted to mitigate the effect of these noisy crowdsourcing labels \cite{zheng2017truth}. Various approaches have been proposed to model the quality \cite{liu2012variational,aydin2014crowdsourcing}, confidence \cite{joglekar2013evaluating}, expertise \cite{ma2015faitcrowd,zheng2016docs}, reliability \cite{li2019modelling} of annotators; or model the difficulty of the tasks \cite{whitehill2009whose,ma2015faitcrowd}. With such information, we can infer the truth label from the noisy labels more accurately and correspondingly train a more desirable model.

In terms of annotator modeling, existing studies mainly concentrated on factors like quality, confidence, expertise, etc., which could affect the annotation results. Besides, the bias or stereotypes held by the annotators can also lead to defective annotations \cite{sap2019risk}, which is, however, rarely studied. In addition, the majority of these studies merely focused on the factors on each individual annotator, but didn't consider the possible group effect in annotators. However, we observe that annotators in different demographic groups tend to show different bias in annotation tasks. For example, in a preliminary study, we examine the instances annotated by both two groups of annotators in the Wikipedia Toxicity dataset~\cite{wulczyn2017ex}. We observe that native speakers of English rate $5.1\%$ more comments as toxic than non-native speakers. Similarly, annotators over 30 years old rate $2.5\%$ more comments as toxic than younger annotators. More details of the preliminary study can be found in Section \ref{sec:understand}. Thus, a thorough investigation on such annotator group bias is desired. Similar to existing studies, by considering the effect of annotator group bias, we have the potential to achieve a more accurate inference of true labels and train a better model. Meanwhile, it is often hard to estimate the individual bias of one annotator with limited annotation data. With annotator group bias as the prior knowledge, we can estimate the bias more effectively based on the demographic groups the annotator belongs to. Thus, annotator group bias could mitigate the ``cold-start'' problem in modeling the annotator individual bias. 



In this paper, we aim to study how to detect annotator group bias under text classification tasks, and how to mitigate the detrimental effects of annotator group bias on model training. We face several challenges. First, given noisy annotated data without the true labels, how should we detect the annotator bias? We first make a comparison of the annotation results from different groups of annotators and find that there is a significant gap between them. Then, we use two metrics \textit{sensitivity} and \textit{specificity} to measure the annotator bias, and conduct an analysis of variance (ANOVA) which demonstrates that the bias of each individual annotator shows obvious group effects in terms of its demographic attributes. Second, how can we estimate the annotator group bias, and perform label aggregation and model training with the knowledge of annotator group bias? Following the traditional probabilistic approaches for label aggregation \cite{raykar2010learning,rodrigues2018deep,li2019modelling}, we propose a novel framework \textbf{GroupAnno} that models the production of annotations as a stochastic process via a novel probabilistic graphical model (PGM). Inspired by the results of ANOVA, we assume that the bias of an individual annotator follows a Beta distribution with a prior subject to its demographic attributes. We thereby extend the original PGM for label aggregation with additional variables representing annotator group bias. By learning the PGM, we estimate the annotator group bias, infer the true labels, and optimize our classification model simultaneously. Third, how can we learn this PGM effectively? With the unknown true label as the latent variable, typical label aggregation models are learned by the Expectation Maximization (EM) algorithm. However, in our PGM, we introduce another latent variable to model the bias of each annotator where the vanilla EM algorithm is not applicable. To address this challenge, we propose an extended EM algorithm for GroupAnno.

We summarize our contributions in this paper as follows. First, we propose metrics to measure the annotator group bias and verify its existence in real NLP datasets via an empirical study. Second, we propose a novel framework GroupAnno to model the annotation process by considering the annotator group bias. Third, we propose a novel extended EM algorithm for GroupAnno where we estimate the annotator group bias, infer the true labels, and optimize the text classification model simultaneously. Finally, we conduct experiments on synthetic and real data. The experimental results show that GroupAnno can accurately estimate the annotator group bias. Also, compared with competitive baselines, GroupAnno can infer the true label more accurately, and learn better classification models.




\vspace{-8pt}
\section{Understanding Annotator Group Bias}
\label{sec:understand}
In this section, we perform an empirical study to get a rudimentary understanding of annotator group bias. 


\subsection{Data and Tasks}
\label{sec:data}
We investigate the group annotator bias on three datasets that involve various text classification tasks. These datasets are released in the Wikipedia Detox project \cite{wulczyn2017ex}: Personal Attack Corpus, Aggression Corpus, and Toxicity Corpus where each instance is labeled by multiple annotators from the Crowdflower platform \footnote{\url{https://www.crowdflower.com/}}.
For all the datasets, the demographic attributes of the annotators are collected. The statistics of these datasets can be found in Appendix A. 




The Personal Attack dataset and the Aggression dataset contain the same comments collected from English Wikipedia. Each comment is labeled by around 10 annotators on two tasks, respectively. The task of the former dataset is to determine whether the comment contains any form of personal attack, while the task of the latter dataset is to judge whether the comment is aggressive or not. For each annotator, four demographic categories are collected: \textit{gender}, \textit{age}, \textit{language}, and  \textit{education}. Although the original dataset provides more fine-grained partitions, for simplicity, we divide the annotators into only two groups in terms of each demographic category. We consider two groups: male and female for \textit{gender}, under 30 and over 30 for \textit{age}, below bachelor and above bachelor (including bachelor) for \textit{education}, and native and non-native speaker of English for \textit{language}. The toxicity dataset contains comments collected from the same source. Similarly, each comment is labeled by around 10 annotators on whether it is toxic or not. The toxicity dataset includes the same demographic information of the annotators as the former two datasets.

\subsection{Empirical Study}
\label{sec:anova}

\begin{table*}[h]
\small
\centering
\caption{The positive rates of the annotations from different groups of annotators.}
\label{tab:compare}
\begin{tabular}{ccccccccc}
\hline
\hline
\multirow{2}{1cm}{\textbf{Dataset}} & \multicolumn{2}{c}{\textbf{\begin{tabular}[c]{@{}c@{}}Gender \end{tabular}}} & \multicolumn{2}{c}{\textbf{\begin{tabular}[c]{@{}c@{}}Age \end{tabular}}} & \multicolumn{2}{c}{\textbf{\begin{tabular}[c]{@{}c@{}}Education \end{tabular}}} & \multicolumn{2}{c}{\textbf{\begin{tabular}[c]{@{}c@{}}Language \end{tabular}}} \\ \cmidrule(r){2-3}  \cmidrule(r){4-5} \cmidrule(r){6-7} \cmidrule(r){8-9}
 & \textbf{Male} & \textbf{Female} & \textbf{Under 30} & \textbf{Over 30} & \textbf{Below Ba.} & \textbf{Above Ba.} & \textbf{Native} & \textbf{Non-native} \\ \hline
\textbf{PersonalAttack}  & 15.98 & 18.67 & 15.83 & 18.52 & 17.63 & 15.81 & 19.95 & 14.40 \\
\textbf{Aggression}  & 17.74 & 21.44 & 17.79 & 20.85 & 20.28 & 17.62 & 23.20 & 16.08 \\
 \textbf{Toxicity}  & 12.06 & 16.37 & 12.51 & 15.08 & 15.16 & 12.56 & 16.93 & 11.80 \\
\hline
\hline

\end{tabular}
\end{table*}

To investigate whether the annotators from different groups behave differently in annotation tasks, we first perform a comparison of the annotation results from different annotator groups. For each demographic category, we collect the instances which are labeled by annotators from both groups, and report the proportion of instances that are classified as positive. The results are shown in Table \ref{tab:compare}. First, we note that there are obvious gaps between the annotations given by different annotator groups. Second, given that the tasks of the three datasets are similar (i.e., all of them are related to detecting inappropriate speech), the annotation tendency of each annotator group is the same. For example, young and non-native speaker annotators are less likely to annotate a comment as attacking, aggressive, or toxic. Third, in terms of different demographic categories, the gaps between the annotations from the two groups are different. For example, compared with other group pairs, the annotations provided by native speakers and non-native speakers are more different. 

\begin{table*}[h]
\small
\centering
\caption{The results of analysis of variance. The table shows the inter-group sum of squares (variance of treatments). *, ** indicate that the group effects are significant at $p<0.05$ and $p<0.005$.}
\label{tab:anova}
\begin{tabular}{cllllll}
\hline
\hline
\multirow{2}{1cm}{\textbf{Category}} & \multicolumn{2}{c}{\textbf{\begin{tabular}[c]{@{}c@{}}Personal Attack \end{tabular}}} & \multicolumn{2}{c}{\textbf{\begin{tabular}[c]{@{}c@{}}Aggression \end{tabular}}} & \multicolumn{2}{c}{\textbf{\begin{tabular}[c]{@{}c@{}}Toxicity \end{tabular}}} \\ \cmidrule(r){2-3}  \cmidrule(r){4-5} \cmidrule(r){6-7}
 & \textbf{Sensitivity} & \textbf{Specificity} & \textbf{Sensitivity} & \textbf{Specificity} & \textbf{Sensitivity} & \textbf{Specificity.} \\ \hline
\textbf{Gender}  & 0.010 & 0.077* & 0.106 & 0.182** & 0.217** & 0.266** \\
\textbf{Age}  & 3.093** & 0.257** & 3.529** & 0.348** & 3.230** & 0.348** \\
\textbf{Education}  & 0.006 & 0.001 & 0.021 & 0.012 & 0.012 & 0.013 \\
\textbf{Language}  & 0.805** & 0.155** & 1.200** & 0.470** & 0.041 & 0.023* \\
\hline
\hline

\end{tabular}
\end{table*}

\textbf{Analysis of Variance.} The results in Table \ref{tab:compare} suggest that annotators show group bias in the annotation tasks, which is manifested in that different groups hold different evaluation criteria in the same task. Specifically for classification tasks, different annotators are unevenly likely to label instances belonging from one class to another class. In this paper, we only consider binary classification tasks for simplicity \footnote{All our findings and the proposed framework can be trivially extended to the case of multi-class classification.}. Thus, we use \textit{sensitivity} (true positive rate) and \textit{specificity} (1 $-$ false positive rate) \cite{yerushalmy1947statistical} to describe the bias of an individual annotator. 

Next, we seek to verify the existence of annotator group bias. We are interested in whether the demographic category of an individual annotator has a significant impact on its bias. Thus, we first estimate the bias (i.e., sensitivity and specificity) of each individual annotator from its annotation data. Since we don't have the true labels, we use majority vote labels as the true labels to approximately estimate the bias of each annotator. Then, we perform an ANOVA \cite{scheffe1999analysis} with the demographic category as the factors, the groups as the treatments, and the bias of an annotator as the response variable, to analyze the significance of the annotator's demographic groups against its own bias. The corresponding statistical model can be expressed as:
\begin{align}
    \Tilde{\pi}_r=u+\pi^{1,g^1_r}+\cdots+\pi^{P,g^P_r}+\epsilon_r
    \label{eq:anova}
\end{align}
\noindent where $\Tilde{\pi}_r$ indicates the bias of an individual annotator $r$; $u$ is the average bias of all annotators; $\pi^{p,g^p_r}$ is the effect of the group $g^p_r$ in terms of category $p$; and $\epsilon_r$ is the random error which follows a normal distribution with the mean value as 0. To test whether category $p$ has a significant impact on $\Tilde{\pi}$, we consider the null hypothesis $H_{0p}: \pi^{p,0}=\pi^{p,1}$, which indicates that the demographic category $p$ has no significant effect on the annotator bias. In other words, there is no significant difference between the annotation behaviors of the two groups in terms of category $p$.%

The results are shown in Table \ref{tab:anova}. In the table, we report the inter-group sum of squares, which represent the deviation of the average group bias from the overall average bias. We also use ``$*$'' to denote the significance of the hypothesis tests. We observe that in categories of gender, age and language, the two opposing groups show obvious different sensitivity and specificity in most cases. Moreover, the ANOVA suggests that we are confident to reject the null hypotheses in these cases, which means that the above three demographic categories can affect the annotator bias significantly in different datasets. Based on our observations, we conclude that the demographic attribute of an annotator can have a significant impact on its annotation behavior, and thereby, annotator group bias does exist.
\section{Modeling Annotator Group Bias}
In this section, we discuss our approaches for annotator group bias estimation, as well as bias-aware label aggregation and model training. We first introduce the metrics for measuring annotator group bias, and then present the problem statement. Next, we detail \textbf{GroupAnno}, the probabilistic graphical model for modeling the production of annotations. Finally, we describe our extended EM algorithm for learning the proposed model.

\subsection{Measurements}
To measure the annotator bias in terms of demographic groups, we extend the definitions of sensitivity and specificity to the group scenario. Formally, we define \textit{group sensitivity} and \textit{group specificity} of a group $g$ in terms of category $p$ as follows %
\begin{align}
    \alpha^{p,g}&=Pr(z=1|y=1,g_r^p=g) \nonumber\\
    \beta^{p,g}&=Pr(z=0|y=0,g_r^p=g) \nonumber
\end{align}
\noindent where $y$ is the true label and $z$ is the annotated label. $g_r^p=g$ represents that the annotator $r$ belongs to group $g$ in terms of demographic category $p$.




We use $\pi^p=(\alpha^{p,0}, \alpha^{p,1}, \beta^{p,0}, \beta^{p,1})$ to denote the bias parameters of demographic category $p$. The bias parameters of all the $P$ categories are denoted as $\pi=\{\pi^p\}_{p=1}^P$.



\subsection{Problem Statement}
Suppose that we have a dataset $\mathbf{D}=\{x_i, z_i^1, \cdots, z_i^{R_i}\}_{i=1}^N$ which contains $N$ instances. Each instance $x_i$ is annotated by $R_i$ different annotators, which results in labels $z_i^1, \cdots, z_i^{R_i}$. We also have an annotator set $\mathbf{A}=\{(g_r^1, \cdots, g_r^P)\}_{r=1}^R$ that records the demographic groups of a total of $R$ annotators. Here, $g_r^p \in \{0,1\}$ indicates the group that the $r$-th annotator belongs to in terms of the $p$-th demographic category. We consider $P$ demographic categories for each annotator, and we have two groups (i.e., 0 and 1) for each category. Given $\mathbf{D}$ and $\mathbf{A}$, we seek to (1) estimate the annotator group bias $\pi$; (2) estimate the true label $y_i$ of each instance $x_i$; and (3) learn a classifier $P_{\mathbf{w}}(y|x)$ which is parameterized by $\mathbf{w}$.

Next, we introduce our GroupAnno to model the annotation process, and propose an extended EM algorithm to estimate the parameters $\Theta=\{\mathbf{w},\pi\}$.


\subsection{GroupAnno: The Probabilistic Graphical Model}

\begin{figure}[h]
\includegraphics[width=\linewidth]{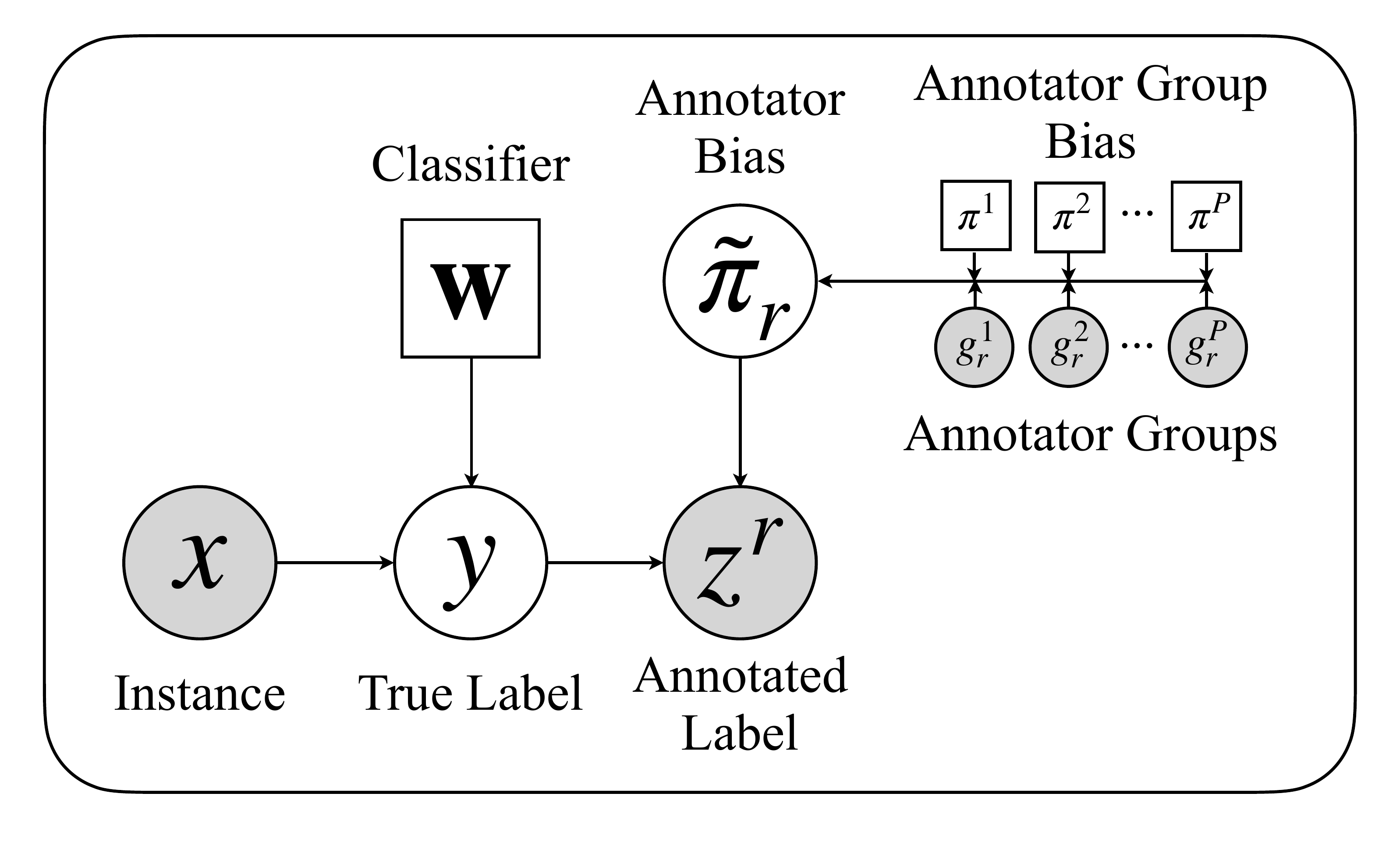}
\caption{An illustration of GroupAnno. In the graph, grey circles represent observed data; white circles indicate latent variables; and squares denote the unknown parameters that we will learn.}
\label{fig:PGM}
\end{figure}

As shown in Figure \ref{fig:PGM}, GroupAnno models the generation procedure of annotations as follows. Given an instance $x$, its true label $y$ is determined by an underlying distribution $P_{\mathbf{w}}(\cdot|x)$. The distribution is expressed via a classifier with parameters $\mathbf{w}$ that we will learn. Given the true label $y$, the annotated label $z^r$ from an annotator $r$ is determined by its bias $\Tilde{\pi}_r=(\Tilde{\alpha}_r, \Tilde{\beta}_r)$ For simplicity, in the following formulations, we use $\Tilde{\pi}_r$ to represent $\Tilde{\alpha}_r$ or $\Tilde{\beta}_r$. To incorporate the prior knowledge from the demographic attributes of an annotator, we assume that the annotator bias $\Tilde{\pi}_r$ follows a Beta prior, which can naturally model the distribution of the probability of a binary event. In Section \ref{sec:anova}, we show that the annotator bias can be modeled by a superposition of the effects of annotator group bias with a random variable reflecting the annotator individual bias. Thus, following Eq \ref{eq:anova}, we assume that the annotator bias of annotator $r$ follows a Beta distribution $\Tilde{\pi}^r \sim Beta(a^r_1, a^r_2)$, where the mean value $\mu_r$ is decomposed as
\begin{align}
    \mu_r = \frac{a^r_1}{a^r_1 + a^r_2} = u+\pi^{1,g^1_r}+\cdots+\pi^{P,g^P_r}+\pi_r \nonumber
\end{align}



To sum up, the parameters we introduced to model annotator bias are $\mathbf{\pi}=\{u\} \cup \{\pi^p\}_{p=1}^P \cup \{\pi_r\}_{r=1}^R$. To estimate the parameters $\Theta=\{\mathbf{w},\pi\}$, one way is to use maximum likelihood estimation (MLE). Under the assumption that instances are sampled independently, the likelihood function of $\Theta$ can be written as
\vspace{-1cm}

\begin{align}
    P(\mathbf{D}|\mathbf{\Theta}) &= \prod_{i=1}^N P(z_i^1, \cdots, z_i^{R_i}|x_i;\Theta)\nonumber
\end{align}

Therefore, the MLE parameters can be found by maximizing the log-likelihood
\begin{align}
    \hat{\Theta}_{MLE}=\{\hat{\mathbf{w}},\hat{\pi}\}= {\rm argmax}_{\Theta} \ln P(\mathbf{D}|\Theta)
    \label{eq:mle}
\end{align}



\subsection{The extended EM algorithm}


Two obstacles determine that we cannot directly apply MLE to solve Equation \ref{eq:mle}. First, in GroupAnno, we suppose that the annotator bias $\Tilde{\pi}_r$ is a random variable with Beta priors, instead of a constant value, which makes traditional MLE method infeasible. Second, there is an unknown latent variable (i.e. the true label $y$) in GroupAnno. Thus, we propose a novel extended EM algorithm to effectively estimate the parameters $\Theta$ in GroupAnno.

To solve the first obstacle, we choose to use a maximum-a-posteriori (MAP) estimator for the optimization, where the prior distribution of the parameters can be considered
\begin{align}
    \hat{\Theta}_{MAP}= {\rm argmax}_{\Theta} \ln P(\mathbf{D}|\Theta) + \ln P(\Theta) \nonumber
\end{align}

Since the true label $y_i$ is an unknown latent variable, the first log-likelihood term can be decomposed as
\begin{align}
    &\ln P(\mathbf{D}|\Theta) \nonumber \\
    &= \sum_{i=1}^N \ln [P_{\mathbf{w}}(y_i=1|x_i) P(z_i^1, \cdots, z_i^{R_i}|y_i=1;\Tilde{\alpha}) \nonumber\\
    & + P_{\mathbf{w}}(y_i=0|x_i) P(z_i^1, \cdots, z_i^{R_i}|y_i=0;\Tilde{\beta})] \nonumber
\end{align}

\noindent where $\Tilde{\alpha}=\{\Tilde{\alpha}_r\}_{r=1}^{R}$ and $\Tilde{\beta}=\{\Tilde{\beta}_r\}_{r=1}^{R}$ represent the collections of the sensitivity and the specificity of all the annotators. We further assume that the annotations for one instance from different annotators are independent \cite{raykar2010learning}. Then we have
\vspace{-10pt}
\begin{align}
    &\ln P(\mathbf{D}|\Theta) \nonumber\\
    &= \sum_{i=1}^N \ln \Big[ P_{\mathbf{w}}(y_i=1|x_i) \times \prod_{r=1}^{R_i} P(z_i^r|y_i=1;\Tilde{\alpha}) \nonumber\\
    & + P_{\mathbf{w}}(y_i=0|x_i) \times \prod_{r=1}^{R_i} P(z_i^r|y_i=0;\Tilde{\beta}) \Big]\nonumber\\
    &= \sum_{i=1}^N \ln [p_i a_i + (1-p_i) b_i]
    \label{eq:lnP}
\end{align}

\noindent where we denote
\begin{small}
\begin{align}
    p_i &:= P_{\mathbf{w}}(y_i=1|x_i)\nonumber\\
    a_i &:= \prod_{r=1}^{R_i} P(z_i^r|y_i=1;\Tilde{\alpha}) = \prod_{r=1}^{R_i} \Tilde{\alpha}_r^{z_i^r} (1-\Tilde{\alpha}_r)^{1-z_i^r}\nonumber\\
    b_i &:= \prod_{r=1}^{R_i} P(z_i^r|y_i=0;\Tilde{\beta}) = \prod_{r=1}^{R_i} (1-\Tilde{\beta}_r)^{z_i^r} \Tilde{\beta}_r^{1-z_i^r}\nonumber
\end{align}
\end{small}

Note that due to the existence of the latent variable $y_i$, Eq \ref{eq:lnP} contains the logarithm of the sum of two terms, which makes it very difficult to calculate its gradient w.r.t $\Theta$. Thus, to solve the second obstacle, we instead optimize a lower bound of $\ln P(\mathbf{D}|\Theta)$ via an EM algorithm.

\textbf{E-step.} Given the observation $\mathbf{D}$ and the current parameters $\Theta$, we calculate the following lower bound of the real likelihood $\ln P(\mathbf{D}|\Theta)$
\begin{align}
    & \ln P(\mathbf{D}|\Theta) \ge \mathbb{E}_\mathbf{y}[\ln P(\mathbf{D}, \mathbf{y}|\Theta)] \ge \nonumber\\
    & \sum_{i=1}^N \mu_i \ln p_i a_i + (1-\mu_i)\ln(1-p_i)b_i + {\rm Const}.
    \label{eq:cond_exp}
\end{align}
\noindent where $\mu_i = P(y_i=1|z_i^1, \dots, z_i^R, x_i, \Theta)$ and it can be computed by the Bayes' rule
\begin{align}
    \mu_i = \frac{a_i p_i}{a_i p_i + b_i (1-p_i)}
    \label{eq:bayes}
\end{align}

\textbf{M-step.} In the M-step, we update the model parameters $\Theta$ by maximizing the conditional expectation in Eq \ref{eq:cond_exp}, as well as its prior probability.
\begin{align}
    \Theta \leftarrow \Theta - \alpha \nabla_\Theta \{\mathbb{E}_\mathbf{y}[\ln P(\mathbf{D},\mathbf{y}|\Theta)]+\ln P(\Theta)\}\nonumber
\end{align}
\noindent where $\alpha$ is the learning rate.

The training algorithm is summarized in Algorithm \ref{alg:HieEM}. We first initialize the posterior probability of the labels $\mu_i$ based on majority voting (line 1). Next, we perform the extended EM algorithm to update the model parameters iteratively. In the E-step, we update $\mu_i$ by Bayes' rule in Eq \ref{eq:bayes}, and then calculate the expectation by Eq \ref{eq:cond_exp} (from lines 3 to 5). Afterward, we perform the M-step, where the gradients of the conditional expectation w.r.t the model parameters are calculated, and the model parameters are updated through gradient descent. The iterative process is terminated when some specific stop requirements are satisfied. In our implementation, we execute the EM optimization steps for a fixed number of epochs.

\begin{algorithm}[h]\small

\KwIn{Dataset $\mathbf{D}=\{x_i, z_i^1, \cdots, z_i^{R_i}\}_{i=1}^N$, annotator set $\mathbf{A}=\{(g_r^1, \cdots, g_r^P)\}_{r=1}^R$.}
\KwOut{a text classification model $\mathbf{w}$, estimated annotator bias $\pi$}

Initialize $\mu_i=\frac{1}{R_i} \sum_{r=1}^{R_i} z_i^r$ based on majority voting.

\Repeat{meets stop requirements}{

\textbf{E-step:} 

Update $\mu_i$: $\mu_i \leftarrow \frac{a_i p_i}{a_i p_i + b_i (1-p_i)}$

Calculate the expectation $\mathbb{E}_\mathbf{y}[\ln P(\mathbf{D},\mathbf{y}|\Theta)+\ln P(\Theta)]$

\textbf{M-step:}

Update the parameters $\Theta$ by maximizing the above expectation.

$\Theta \leftarrow \Theta - \alpha \nabla_\Theta \{\mathbb{E}_\mathbf{y}[\ln P(\mathbf{D},\mathbf{y}|\Theta)]+\ln P(\Theta)\}$

}
    
\caption{{\bf The optimization algorithm.} \label{alg:HieEM}}

\end{algorithm}
\section{Experiment}
In this section, we evaluate the proposed method via comprehensive experiments. We test our model on both synthetic and real-world data. Through the experiments, we try to answer three research questions: (1) is our method able to accurately estimate the annotator group bias? (2) can our method effectively infer the true labels? and (3) can our approach learn more accurate classifiers?

\subsection{Baselines}
We compare our proposed framework GroupAnno with nine existing true label inference methods \cite{zheng2017truth}, including majority voting (MV), ZenCrowd \cite{demartini2012zencrowd}, Minimax \cite{zhou2012learning}, LFC-binary \cite{raykar2010learning}, CATD \cite{li2014confidence}, PM-CRH \cite{aydin2014crowdsourcing}, KOS \cite{karger2011iterative}, VI-BP and VI-MF \cite{liu2012variational}.

\subsection{Data}

\begin{figure}[!t]
\includegraphics[width=\linewidth]{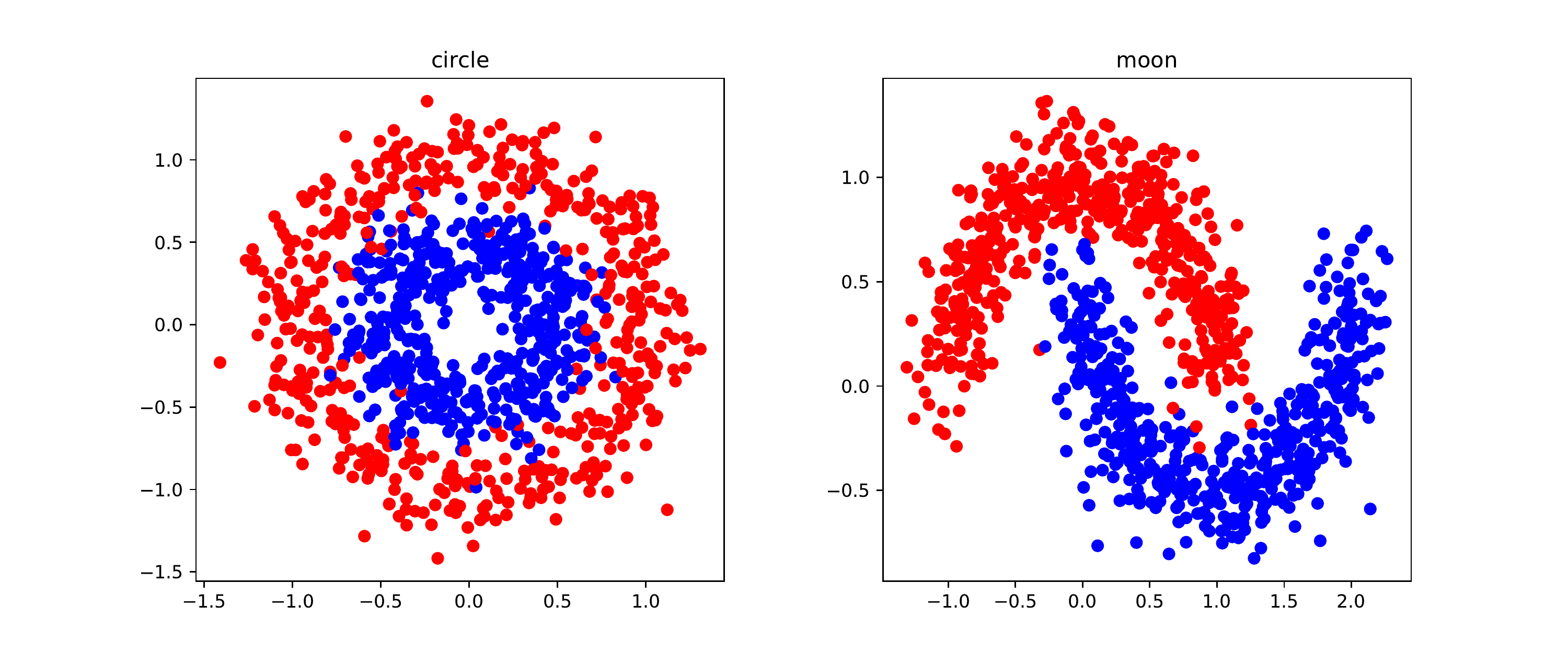}
\caption{Two synthetic datasets with simulated 2-dimensional data.}
\label{fig:2dim}
\end{figure}

\textbf{Synthetic Data.} We first create two synthetic datasets on a simple binary classification task with 2-dimension features. As shown in Figure \ref{fig:2dim}, the instances in the datasets are in the shape of circle and moon, respectively. In each dataset, we sample 400 instances for both classes. We simulate 40 annotators with two demographic attributes. We suppose that each instance is labeled by 4 different annotators, and simulate the annotations based on pre-defined annotator bias. With the knowledge of actual annotator group bias and true labels in synthetic data, we can verify the capability of the proposed framework in group bias estimation and truth label inference.

\textbf{Wikipedia Detox Data.} We conduct experiments on all the three subsets (i.e. Personal Attack, Aggression, and Toxicity) of the public Wikipedia Detox dataset. The details of this dataset are introduced in Section \ref{sec:data}. More details about the experimental setup on this dataset can be found in Appendix A.2.

\textbf{Information Detection Data.} This dataset consists of text transcribed from conversations recorded in several in-person and virtual meetings. Each text is assigned a information label which groups the text into three categories: give information (G), ask information (A), and other (O). Five different data annotators classified the text into one of G, A, or O categories. We conducted a survey to collect data on demographic characteristics of the annotators such as gender, race, and native speaker of English. We convert the three categories into two classes by treating G and A as positive (i.e., information exchange) and O as negative (i.e., other). More details about this dataset can be found in Appendix B. 



\subsection{Results on Synthetic Data}

\textbf{Group Bias Estimation.} In each of the synthetic datasets, we simulate the annotations based on presented annotator group bias. We simulate two demographic attributes for each annotator, where there are two groups in terms of each attribute. Thus, there are eight bias parameters to estimate: sensitivities $\alpha^{p,g}$ and specificities $\beta^{p,g}$, where $p=0,1$ and $q=0,1$. We compare the real values of the annotator group bias and the estimations from GroupAnno. The results are shown in Table \ref{tab:bias_est}. We observe that the bias parameters are estimated accurately within an acceptable error range. The results demonstrate the ability of our extended EM algorithm to estimate the parameters in GroupAnno.

\begin{table}[t]
\small
\centering
\caption{Results of group bias estimation on the synthetic 2-dimensional datasets. ``Real'' and ``Estimation'' indicate the real and the estimated values of the annotator group bias parameters.}
\vspace{8pt}
\label{tab:bias_est}
\begin{tabular}{cccc}
\hline
\hline
\multirow{2}{1cm}{\textbf{Params}} & \multirow{2}{1cm}{\textbf{Real}} & \multicolumn{2}{c}{\textbf{\begin{tabular}[c]{@{}c@{}}Estimation \end{tabular}}} \\  \cmidrule(r){3-4}
 & & \textbf{Circle} & \textbf{Moon}\\ \hline
$\alpha^{0,0}$ & 0.700 & 0.739 & 0.728\\
$\alpha^{0,1}$ & 0.500 & 0.482 & 0.476\\
$\beta^{0,0}$ & 0.800 & 0.787 & 0.778\\
$\beta^{0,1}$ & 0.300 & 0.335 & 0.320\\
\hline
$\alpha^{1,0}$ & 0.900 & 0.927 & 0.943\\
$\alpha^{1,1}$ & 0.400 & 0.419 & 0.428\\
$\beta^{1,0}$ & 0.300 & 0.288 & 0.295\\
$\beta^{1,1}$ & 0.500 & 0.458 & 0.443\\
\hline
\label{tab:syn_result}
\end{tabular}
\vspace{-10pt}
\end{table}

\textbf{Truth Label Inference.} The experimental results of truth label inference on synthetic data are shown in Table \ref{tab:syn_result}. In the table, we list the performance of different approaches on truth label inference. We make the following observations. First, among the baselines, Minimax, KOS, VI-BP, and VI-MF perform significantly worse than others. The reasons are: (1) Minimax and KOS don't model the sensitivity and specificity of an annotator but only use one single value to model the quality of an annotator. In fact, the labeling accuracy of an annotator can be very different when the ground-truth label is positive and negative. Hence, only modeling a single annotator quality parameter can lose important information. (2) Although VI-BP and VI-MF consider the sensitivity and specificity, they exploit variational inference to estimate these parameters, which has been shown to be an ineffective approach on this task \cite{zheng2017truth}. Second, LFC-binary outperforms other baselines. LFC-binary leverages PGM to model the individual annotator bias for truth label inference, which achieves desirable performance. Third, our framework GroupAnno further improves the accuracy of truth label inference on the basis of LFC-binary, since GroupAnno finds and exploits the group annotator bias as additional information. GroupAnno models the group annotator bias as prior information of the individual bias of each annotator so that individual bias can be estimated more accurately. As a result, GroupAnno achieves the best performance on truth label inference.


\begin{table}[t]
\small
\centering
\caption{Experimental results on the synthetic 2-dimensional datasets. ``Acc'' and ``F1'' indicate the accuracy and the F1 score of true label inference.}
\vspace{8pt}
\label{tab:2-dim}
\begin{tabular}{ccccc}
\hline
\hline
\multirow{2}{1cm}{\textbf{Methods}} & \multicolumn{2}{c}{\textbf{\begin{tabular}[c]{@{}c@{}}Circle \end{tabular}}} & \multicolumn{2}{c}{\textbf{\begin{tabular}[c]{@{}c@{}}Moon \end{tabular}}} \\ \cmidrule(r){2-3}  \cmidrule(r){4-5}
 & \textbf{Acc} & \textbf{F1} & \textbf{Acc} & \textbf{F1} \\ \hline
\textbf{MV} & 0.716 & 0.724 & 0.751 & 0.731 \\
\textbf{ZenCrowd} & 0.894 & 0.886 & 0.904 & 0.898 \\
\textbf{Minimax} & 0.499 & 0.490 & 0.516 & 0.529 \\
\textbf{LFC-binary} & 0.911 & 0.909 & 0.916 & 0.914 \\
\textbf{CATD} & 0.850 & 0.843 & 0.858 & 0.850 \\
\textbf{PM-CRH} & 0.861 & 0.853 & 0.879 & 0.872 \\
\textbf{KOS} & 0.502 & 0.519 & 0.511 & 0.501 \\
\textbf{VI-BP} & 0.505 & 0.543 & 0.475 & 0.484 \\
\textbf{VI-MF} & 0.504 & 0.496 & 0.484 & 0.479 \\
\textbf{GroupAnno} & \textbf{0.925} & \textbf{0.924} & \textbf{0.926} & \textbf{0.925}\\
\hline
\hline
\label{tab:syn_result}
\vspace{-10pt}
\end{tabular}
\end{table}

\subsection{Results on Wikipedia Detox Dataset}


The experimental results on the Wikipedia Detox datasets are shown in the left section of Table \ref{tab:wiki}. For LFC-binary and GroupAnno, where truth label inference and model training are conducted simultaneously, we directly report the performance of the resulting model on the test set. For other pure truth label inference approaches, we first infer the truth labels and then train the model on the inferred labels. Finally, we report the performances of these models on the test set. The results show that GroupAnno improves the state-of-the-art methods by a significant margin, which demonstrates the effectiveness and superiority of our framework in practice.

\begin{table*}[t]
\small
\centering
\caption{Expermental results on the Wikipedia Detox datasets and the Information Detection dataset. For Wikipedia Detox, we report the performances of the learned classifiers on the test data. For Information Detection, we report the performance on truth inference (``Truth Infer'') as well as the performance of the learned classifiers on the test data (``Prediction'').}
\vspace{8pt}
\label{tab:wiki}
\begin{tabular}{c|cccccc|cccc}
\hline
\hline
\textbf{Dataset} & \multicolumn{6}{c|}{\textbf{Wikipedia Detox}} & \multicolumn{4}{c}{\textbf{Information Detection}}\\ 
\hline
\multirow{2}{1cm}{\textbf{Method}} & \multicolumn{2}{c}{\textbf{Aggression}} & \multicolumn{2}{c}{\textbf{Personal Attack}} & \multicolumn{2}{c|}{\textbf{Toxicity}} & \multicolumn{2}{c}{\textbf{Truth Infer}} & \multicolumn{2}{c}{\textbf{Prediction}} \\
 & \textbf{Acc} & \textbf{F1} & \textbf{Acc} & \textbf{F1} & \textbf{Acc} & \textbf{F1} & \textbf{Acc} & \textbf{F1} & \textbf{Acc} & \textbf{F1}\\ \hline 
\textbf{MV} & 0.803 & 0.513 & 0.779 & 0.459 & 0.791 & 0.430 & 0.786 & 0.862 & 0.846 & 0.901 \\
\textbf{ZenCrowd} & 0.947 & 0.830 & 0.950 & 0.815 & 0.950 & 0.802 & 0.786 & 0.862 & 0.846 & 0.901 \\
\textbf{Minimax} & 0.502 & 0.246 & 0.501 & 0.224 & 0.519 & 0.190 & 0.823 & 0.872 & 0.855 & 0.898\\
\textbf{LFC-binary} & 0.959 & 0.851 & 0.951 & 0.836 & 0.958 & 0.839 & 0.814 & 0.872 & 0.864 & 0.908\\
\textbf{CATD} & 0.928 & 0.767 & 0.926 & 0.752 & 0.930 & 0.737 & 0.793 & 0.848 & 0.835 & 0.877\\
\textbf{PM-CRH} & 0.933 & 0.782 & 0.932 & 0.767 & 0.936 & 0.756 & 0.782 & 0.831 & 0.820 & 0.845 \\
\textbf{KOS} & 0.508 & 0.230 & 0.508 & 0.219 & 0.499 & 0.188 & 0.786 & 0.862 & 0.843 & 0.899\\
\textbf{VI-BP} & 0.503 & 0.233 & 0.501 & 0.215 & 0.503 & 0.190 & 0.247 & 0.000 & 0.241 & 0.000\\
\textbf{VI-MF} & 0.503 & 0.228 & 0.500 & 0.214 & 0.504 & 0.183 & 0.823 & 0.872 & 0.855 & 0.898\\
\textbf{GroupAnno} & \textbf{0.963} & \textbf{0.862} & \textbf{0.960} & \textbf{0.839} & \textbf{0.962} & \textbf{0.842} & \textbf{0.825} & \textbf{0.883} & \textbf{0.869} & \textbf{0.910}\\
\hline
\hline

\end{tabular}
\end{table*}

\subsection{Results on Information Detection Dataset}

The experimental results on the information detection dataset are shown in the right section of Table \ref{tab:wiki}. Since we have 20\% training data with available true labels, we first examine the accuracy of truth label inference of various methods on this part of the data, and then report the performance of the trained classifiers on the test data. We find that our proposed method still outperforms all the baselines by a significant margin on both truth inference and resulting classifier performance, which further verifies the superiority of GroupAnno in real-world data.

\section{Related Work}

Bias and fairness issues are crucial as machine learning systems are being increasingly used in sensitive applications~\cite{chouldechova2018frontiers}.
Bias is caused due to pre-existing societal norms~\cite{friedman1996bias}, data source, data labeling, training algorithms, and post-processing models.
Data source bias emerges when the source distribution differs from the target distribution where the model will be applied~\cite{shah2019predictive}. Training algorithms can also introduce bias. For example, if we train a model on data that contain labels from two populations - a majority and a minority population - minimizing overall error will fit only the majority population ignoring the minority~\cite{chouldechova2018frontiers}. Data labeling bias exists when the distribution of the dependent variable in the data source diverges from the ideal distribution~\cite{shah2019predictive}. Many of these data labels are generated by human annotators, who can easily skew the distribution of training data~\cite{dixon2018measuring}. Various factors such as task difficulty, task ambiguity, amount of contextual information made available, and the expertise of the annotator determine annotation results~\cite{joseph2017constance}.

Prior literature studies various approaches to ensure the reliability of data annotations.
\citet{demartini2012zencrowd,aydin2014crowdsourcing} use worker probability to model the ability of an annotator to correctly answer a task, and some other works \cite{whitehill2009whose,li2014resolving} introduce a similar concept, worker quality, by changing the value range from $[0,1]$ to $(-\infty, +\infty)$. \citet{welinder2010multidimensional} model the bias and variance of the crowdsourcing workers on numeric annotation tasks. Moreover, \citet{fan2015icrowd} and \citet{ma2015faitcrowd} find that annotators show different quality when answering different tasks, and thereby propose to model the diverse skills of annotators on various tasks. \citet{li2019modelling} realize that annotators perform unevenly on each annotation instance, so they propose a novel method to model the instance-level annotator reliability for NLP labeling tasks. \citet{geva2019we} use language generated by annotators to identify annotator identity and showed that annotator identity information improves model performance. All these studies have been individual-focused and ignore group effects. Our approach differs in that we study systemic bias associated with annotators of a specific demographic group.

\section{Conclusion}
In this work, we investigate the annotator group bias in crowdsourcing. We first conduct an empirical study on real-world crowdsourcing datasets and show that annotators from the same demographic groups tend to show similar bias in the annotation tasks. We develop a novel framework GroupAnno that considers the group effect of annotator bias, to model the whole annotation process. To solve the optimization problem of the proposed framework, we propose a novel extended EM algorithm. Finally, we empirically verify our approach on two synthetic datasets and four real-world datasets. The experimental results show that our model can accurately estimate the annotator group bias, achieve more accurate truth inference, and also train better classifiers that outperform those learned under state-of-the-art true label inference baselines. As future work, we plan to investigate the annotator group bias in tasks beyond classification such as regression tasks and text generation tasks.



\bibliographystyle{acl_natbib}
\bibliography{main}




\end{document}